# Effect of the magnetic domain structure in the ferromagnetic contact on spin accumulation in silicon

Y. Ando[1,2,*], S. Yamada[1], K. Kasahara[1], K. Sawano[3], M. Miyao[1], and K. Hamaya[1,4,**]

[1]Department of Electronics, Kyushu University, Fukuoka 819-0395, Japan
[2]INAMORI FRC, Kyushu University, Fukuoka 819-0395, Japan
[3] Research Center for Silicon Nano-Science, Advanced Research Laboratories, Tokyo City University, Tokyo 158-0082, Japan
[4]PRESTO, Japan Science and Technology Agency, Kawaguchi, Saitama 332-0012, Japan

## Abstract

We show a marked effect of the magnetic domain structure in an epitaxial CoFe contact on the spin accumulation signals in Si detected by three-terminal Hanle-effect measurements. Clear reduction in the spin accumulation signals can be seen by introducing the domain walls in the CoFe contact, caused by the lateral spin transport in the Si channel. The domain walls in the CoFe contact largely affect the spin lifetime and bias-current dependence of the spin signals. These results indicate that the estimation of the spin related properties without considering the domain structure in the contact causes non-negligible errors in the three-terminal Hanle-effect measurements.

Keywords: three-terminal Hanle-effect measurement, spin accumulation, domain configuration, spin diffusion

*Corresponding author: E-mail: ando@ee.es.osaka-u.ac.jp
**hamaya@ed.kyushu-u.ac.jp.

Recently, the nonequilibrium spin accumulation in semiconductors (SCs) was detected electrically by using three-terminal Hanle-effect measurements (3T measurements) [1-11]. Spin related properties such as spin injection efficiency from a ferromagnetic metal (FM) into a SC and spin lifetime in the SC were also discussed by analyzing the spin accumulation signals [1-11]. However, reliability of the estimated values of the spin related properties is now open for discussion. For instance, the experimental value of the magnitude of the spin accumulation signal was reported to be 2-6 k$\Omega \mu m^2$ for the Si based devices with highly doped channel (impurity density: ~$10^{19}$ cm$^{-3}$), which is several orders of magnitude larger than the theoretical one (0.001 k$\Omega \mu m^2$) [12]. Considerably short spin lifetime of ~ 0.1 ns, which is less than one-tenth of that estimated by nonlocal techniques, was also reported [12]. Therefore, elucidating the origin of these discrepancies is strongly desired.

Some possible origins have so far been pointed out by several groups. Contribution of other transport processes through the FM/SC interface such as localized states has the potential to modify the magnitude of the spin accumulation signals [2]. The lateral inhomogeneity of the tunnel current density makes it more difficult to estimate the junction size which has been utilized for the theoretical estimation of spin related properties [3, 10]. Interface roughness is also regarded as a cause, which can reduce the amount of spin accumulation and spin lifetime [6]. Unfortunately they are nothing more than speculations.

Hereafter, we focus on the effect of the magnetic domain structure in the FM contact used on the spin accumulation signals as one of the possible origins of the discrepancies between theories and experiments, because variously-sized FM contacts which can give rise to various magnetic domain structures have so far been utilized for the 3T measurements in the previous studies [1-11]. For the spin valve devices with multiple FM contacts which can detect the spin accumulation after intermixing of whole injected spins, it has been revealed that the magnetic domain structure of the spin injector strongly affects the spin accumulation signals [13]. For the 3T measurements which generally utilize a single FM contact, on the other hand, the effect of the magnetic domain structure has so far been regarded to be negligible by assuming that the spin accumulation before intermixing can be detected. In fact, many groups have performed theoretical estimation of the spin related properties without considering the effect of the domain structure in spite of using a large-scale FM contact which cannot form a single domain structure. When the channel depth is considerably small, the injected spins can be transported laterally in the vicinity of the FM/SC interface. As a result, the spin accumulation created at downstream point may be strongly affected by the spins created at upstream point[1, 14]. That is, the effect of the magnetic domain structure on the spin accumulation signals has to be considered in this situation.

In this letter, using a Si-based three-terminal device with an epitaxial CoFe contact whose domain structures at the remanence can be easily controlled, we show an unexpectedly-large effect of the magnetic domain structure on the magnitude of spin accumulation signals, $|\Delta V_{\text{Hanle}}|$, even for the 3T measurements. The magnetic domain structure also strongly affects the estimated spin lifetime and bias-current dependence of the spin accumulation signal. These results clearly indicate that, in the 3T measurements, estimation of the spin related properties without considering the domain structure causes non-negligible errors, and spin injector whose domain structures can completely be controlled should be utilized for analyzing the spin signals.

10-nm-thick $Co_{60}Fe_{40}$ epitaxial layer was grown on Si(111) by low-temperature molecular beam epitaxy (LT-MBE) at 60 °C [15]. The CoFe/Si(111) interface was atomically flat and was utilized for the three-terminal lateral device [4, 7], as shown in Fig. 1(a). The three-terminal device with an $n$-Si channel (channel thickness ~ 100 nm, carrier density at 300 K ~6.0 × $10^{17}$ cm$^{-3}$) was fabricated by using conventional processes with photolithography, Ar$^+$ ion milling, and reactive ion etching [4, 7]. The epitaxial CoFe contact and two AuSb ohmic ones have lateral dimensions of 10 × 200 μm$^2$ and 100 × 200 μm$^2$, respectively. To achieve tunneling conduction through the high-quality CoFe/Si interface, we inserted an Sb δ-doped $n^+$-Si layer (Sb ~5×10$^{19}$ cm$^{-3}$) with a thickness of ~5 nm between the epitaxial CoFe layer and the $n$-Si channel. The 3T measurements were performed by a dc method in the current-voltage terminal configuration shown in Fig. 1(a), where a small magnetic field (~200 Oe) perpendicular to the plane, $B_Z$, was applied to induce spin depolarization in the Si channel. To control domain configuration in the CoFe contact at the remanence, the in-plane magnetic field ($B_{\text{In}}$ = ~ 2000 Oe) was applied along the long ($\theta$ = 0°) or short ($\theta$ = 90°) axis of the contact before the 3T measurements. During the 3T measurements, the $B_{\text{In}}$ was removed and only $B_Z$ was applied.

Firstly, remanent magnetic domain configurations in the CoFe contact are investigated by means of magnetic force microscope (MFM) after applying two different $B_{\text{In}}$, $\theta$ = 0° and 90°. Figures 1(b) and 1(c) show enlarged MFM images of some parts of the epitaxial CoFe contact at the remanence for $\theta$ = 0° and 90°, respectively, at 300 K. For $\theta$ = 0° [Fig. 1(b)], we confirm almost no clear domain wall at several points of the CoFe contact, indicating the formation of the single-domain structure. On the other hand, for $\theta$ = 90° [Fig. 1(c)], several domain walls are observed, clearly indicating the formation of the multi-domain structure. We also confirmed that the remanent state keeps the same images shown in Figs. 1(b) and 1(c) under small perpendecular magnetic fields below 300 Oe. Furthermore, anisotropic magneto-resistance data at low temperatures can guarantee the same domain configurations, as shown in Figs. 1(b) and 1(c), even for the low temperatures. As a consequence, we can easily control the magnetic domain structure of the CoFe contact during 3T measurements by using $B_{In}$ application process before the measurements.

In Fig. 2(a) we show representative $\Delta V_{3T}$-$B_z$ curves for $\theta = 0°$ (single-domain) and 90° (multi-domain) at $I = 1$ μA at 40 K, where the electrons are extracted from the Si conduction band, i.e., spin-extraction conditions. Quadratic background voltages depending on $B_z$ are subtracted from the raw data. For $\theta = 0°$, $|\Delta V_{Hanle}|$ of ~ 30 μV is clearly observed while that of ~ 21 μV obtained for $\theta = 90°$. That is, by introducing the domain walls in the CoFe contact, $|\Delta V_{Hanle}|$ is markedly reduced from ~ 30 μV to ~ 20 μV. Estimation of the spin lifetime is also performed by using a simple Lorentzian function, $\Delta V_{3T}(B_z) = \Delta V_{3T}(0)/[1+(\omega_L \tau_S)^2]$, where $\omega_L = g\mu_B B_z$ is the Larmor frequency, $g$ is the electron $g$-factor ($g = 2$), $\mu_B$ is the Bohr magneton, and $\tau_S$ is the lower limit of spin lifetime [3]. The fitting results are denoted by the solid curves in Fig. 2(a). The $\tau_S$ value at 40 K for $\theta = 90°$ is estimated to be 1.41 ns, slightly smaller than that for $\theta = 0°$ ($\tau_S = 1.87$ ns). Note that the $\tau_S$ value is also varied despite the use of the same contact at the same temperature. This means that the difference in the magnetic domain structure of the CoFe contact probably affects the lifetime of accumulated spins in the Si channel.

The main panel of Fig. 2(b) shows plots of $|\Delta V_{Hanle}|$ vs. $I$ at 40 K for $\theta = 0°$ and 90°. The $|\Delta V_{Hanle}|$ value for $\theta = 90°$ is obviously smaller than that for $\theta = 0°$ in all $I$ conditions. The effect of the domain structure on the $|\Delta V_{Hanle}|$ value in the spin injection conditions ($I < 0$) is more noticeable than that in the spin-extraction conditions ($I > 0$). For example, we show $\Delta V_{3T}$-$B_z$ curves at $I = -2$ μA in the inset of Fig. 2(b). The $|\Delta V_{Hanle}|$ value for $\theta = 90°$ is estimated to be ~ 3 μV, which is about four times smaller than that for $\theta = 0°$ ($|\Delta V_{Hanle}| = \sim 13$ μV). The estimated $\tau_S$ value is also strongly reduced from 3.56 to 2.68 ns due to the existence of the domain walls. It should be noted that the $|\Delta V_{Hanle}|$ value for $\theta = 90°$ at $I = -2$ μA is obviously smaller than that for $\theta = 0°$ at $I = -0.2$ μA despite the tenfold larger $I$. Here, we focus on the data in a small $I$ region of the spin-injection conditions ($-1.0$ μA $\leqq I < 0$). Both for $\theta = 0°$ and 90°, there are some conditions where $|\Delta V_{Hanle}|$ cannot be detected, as indicated by arrows. These conditions were seen when the energy level of the accumulated spins was lower than the Fermi energy of the CoFe, $E_F^{CoFe}$ [4]. In this condition, the tunneling electrons cannot participate in the spin dependent tunneling, as shown in the right figure of Fig. 2(c). We also note that the range of such $I$ condition for $\theta = 90°$ is relatively wide compared with

that for $\theta = 0°$. In other words, even for the same $I$ conditions, there are somewhat differences in the sensitivity of the spin detection whether the domain structure of the CoFe contact is the single domain or not.

Figure 3(a) schematically shows a possible mechanism of the reductions in $|\Delta V_{\text{Hanle}}|$ and $\tau_S$. Here, we consider the electrochemical potentials in a Si channel for the spin extraction conditions. For $\theta = 90°$ (multi-domain), the spins with various directions are created by the spin extraction to the various magnetic domains, as shown in the right figure of Fig. 3(a). Simultaneously, lateral diffusion of the accumulated spins occurs. As a result, the average of the energy split of the electrochemical potential between up and down spins is reduced compared with that for $\theta = 0°$ (single-domain). If the effect of lateral spin diffusions in the Si channel is considered, $\tau_S$ obtained here can further changed. The actual $\tau_S$ for $\theta = 90°$ should become shorter than that for $\theta = 0°$ due to the additional lateral diffusion of the accumulated spins [16]. If one used large FM contacts including lots of domain walls for 3T measurements, one may do the significant underestimation of the $\tau_S$ value from the simple Lorentzian function. In the spin injection conditions, since the spin accumulation below $E_F^{\text{CoFe}}$ is not detected in the 3T measurements as shown in Fig. 2(c) [4], the reduction in the spin accumulation due to the presence of the domain walls can cause not only reduction in $|\Delta V_{\text{Hanle}}|$ but also the disappearance of the spin accumulation signals in a certain bias condition.

In order to show the validity of considering lateral diffusion of accumulated spins in the Si channel, we analyze the bias-dependent features shown in Fig. 2(c). In general, the bias current dependence of the spin accumulation signals can be numerically calculated as follows.[1]

$$\Delta V(x_1, x_2, B) = \int_0^\infty \frac{V_0}{\sqrt{4\pi D t}} e^{-\left(\frac{(x_2-x_1+v_d t)^2}{4Dt} - \frac{t}{\tau_s}\right)} \times \cos\left(\frac{g\mu_B B}{\hbar} t\right) dt,$$

where $D$ is the diffusion constant, $v_d$ is the drift velocity, $g$ is the electron $g$-factor ($g = 2$), and $\mu_B$ is the Bohr magneton. This equation describes the spin polarization at the detection point $x_2$ at a time $t$ after injection $x_1$ at $t = 0$. To estimate the spin lifetime, we integrated $\Delta V$ in all $x_1$, $x_2$ conditions ($0 \leq x_1, x_2 \leq 10$ μm). The spin lifetime $\tau_s$, contact area of the spin injector, and the diffusion constant $D$ were set to 1.87 ns, 10 × 200 μm$^2$, and 3.45 cm$^2$/s, respectively. The directions of the magnetization and injected spins are also considered as shown in the inset of Fig. 3(b). If we define $\frac{|\Delta V_{\text{Hanle}}|_{\theta=90°}}{|\Delta V_{\text{Hanle}}|_{\theta=0°}}$ as α, a plot of α versus bias current can be calculated, as shown in Fig. 3(b). The α value decreases with increasing the bias current due to enhancement of the spin drift velocity.[1] This feature is in good accordance with that of the experimental data in

Fig.3(c). From these considerations, we verify that the reduction in the $|\Delta V_{\text{Hanle}}|$ due to the multi-domain structure in the FM contact originates dominantly from the lateral diffusion of the accumulated spins in the Si channel.

We also consider the effect of the stray fields from the magnetic domain walls on the spin accumulation signals. In general, the stray fields parallel to the short axis of the CoFe contact (Y axis in Fig. 1(a)) and perpendicular to the plane of the Si channel (Z axis in Fig. 1(a)) can induce the precession of the accumulated spins and modulate the spin accumulation signals [6]. However, to obtain $\alpha = 0.6 \sim 0.7$, typical experimental value in this study, the effective area of the stray fields is roughly estimated to be above 600 μm², which is considerably larger than that observed in MFM images such as Fig. 1(b). Furthermore, whereas the stray fields are expected to give a constant α value irrespective of the bias current, the experimental data in Fig. 3(c) has clear bias current dependence. Therefore, we can conclude that the reduction in the $|\Delta V_{\text{Hanle}}|$ does not originate from the stray fields from the domain walls but from the influence of the lateral spin diffusion in the Si channel.

Finally, we comment on the effect of the sensitivity of the spin detection by using 3T measurements. In our previous studies, we reported that $|\Delta V_{\text{Hanle}}|$ depends strongly on the interface resistance $R_{\text{Int}}$ between CoFe and Si because the sensitivity of the spin detection was changed with changing the $R_{\text{Int}}$ [4]. However, since the $R_{\text{Int}}$ in this study is constant irrespective of the domain structure in the FM contact, there is no critical difference in the sensitivity of spin detection between $\theta = 0°$ (single-domain) and 90° (multi-domain). Thus, we do not have to consider the effect of the change in the sensitivity reported in ref. 4.

In summary, we have investigated the influence of the magnetic domain structure in the CoFe contact on the spin-accumulation signals in 3T measurements. We observed a marked reduction in the spin accumulation signal by introducing the domain walls in the CoFe contact. This feature can be understood by the influence of the lateral spin diffusion in the Si channel. For precisely understanding physical properties in Si spintronics [17-21], it is quite important to take into account the control of the magnetic domain structure in a spin injector and detector.

This work was partly supported by PRESTO-JST, STARC, and SCOPE from MIC, Japan.

**Figure Captions**

Fig.1 (a) Schematic diagram of a lateral three-terminal device using a CoFe/Si contact. MFM images of the CoFe contact measured at 300 K for (b) $\theta = 0°$ and (c) $90°$, respectively.

Fig.2 (a) $\Delta V_{3T}$-$B_Z$ curves for $I = 1$ μA at 40 K for $\theta = 0°$ (circle) and $90°$ (triangle), respectively. (b) $|\Delta V_{Hanle}|$ as a function of bias-current $I$ at 40 K. The inset shows $\Delta V_{3T}$-$B_Z$ curves for $I = -2$ μA. (c) Schematic diagrams of the spin accumulation in the spin injection conditions.

Fig.3 (a) Schematic illustrations of possible electrochemical potentials in a Si channel in the spin extraction conditions. (b) Calculated α value as a function of bias current. (c) Bias current dependence of the α value for spin extraction conditions measured at 40 K.

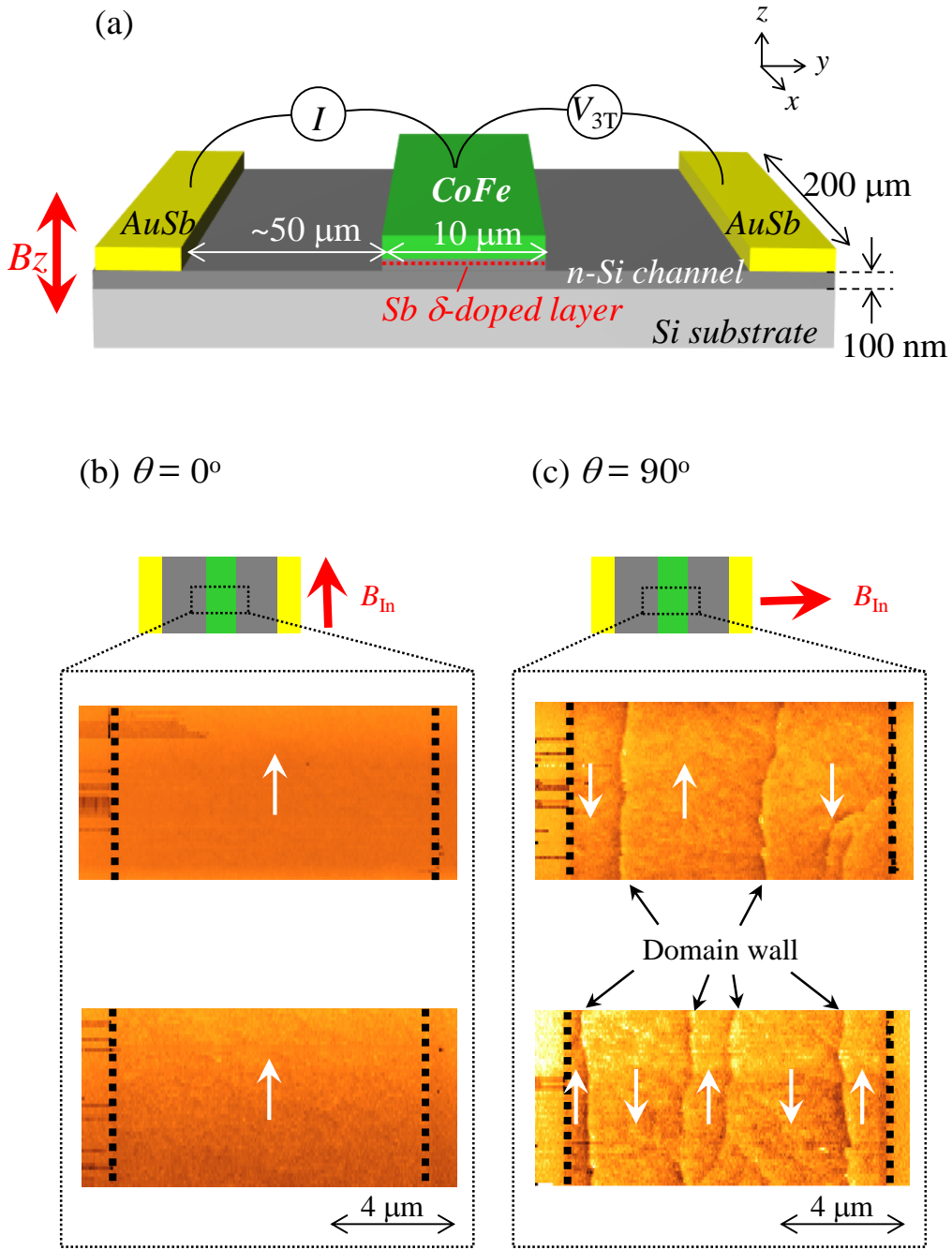

Fig. 1 Y. Ando

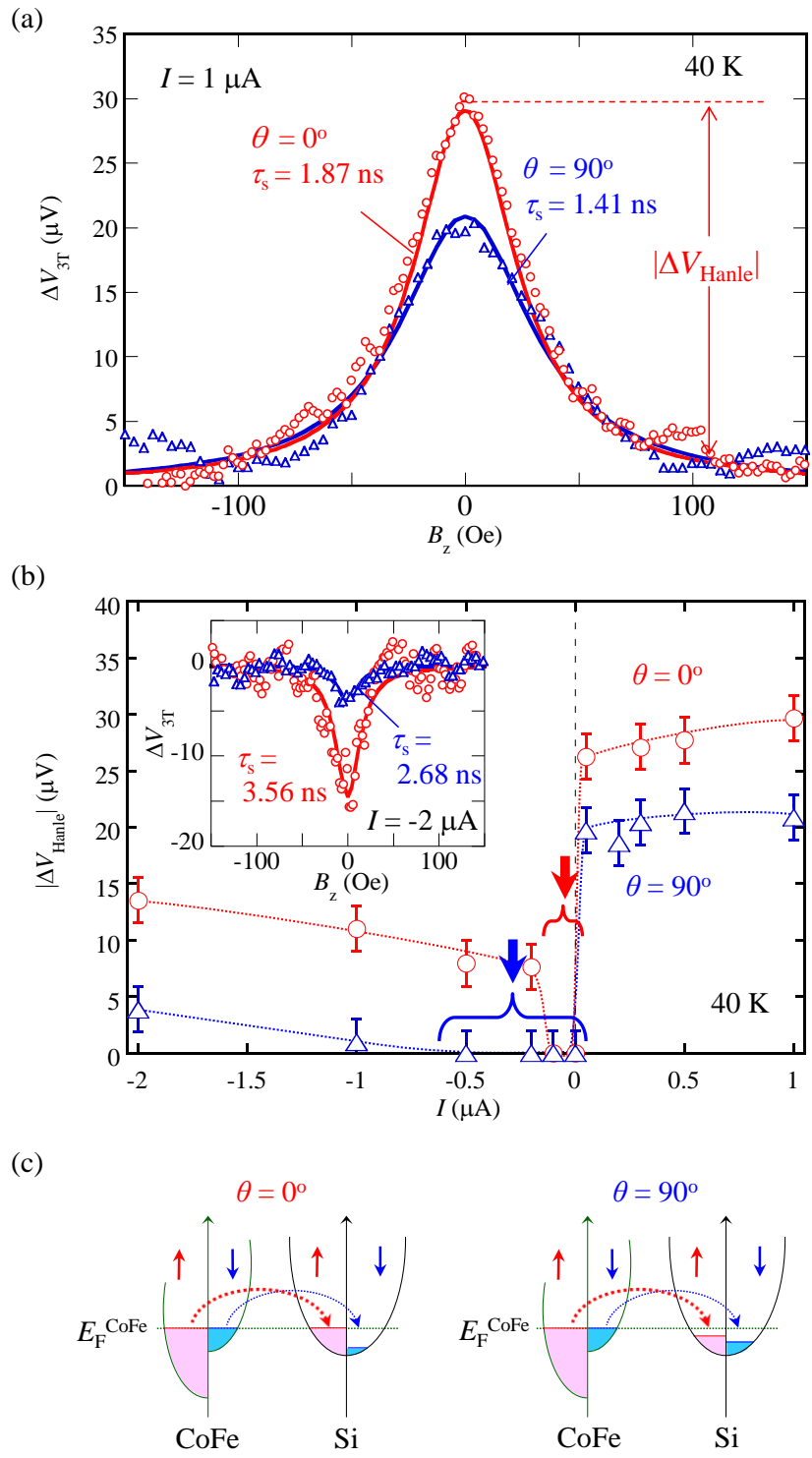

Fig. 2 Y. Ando

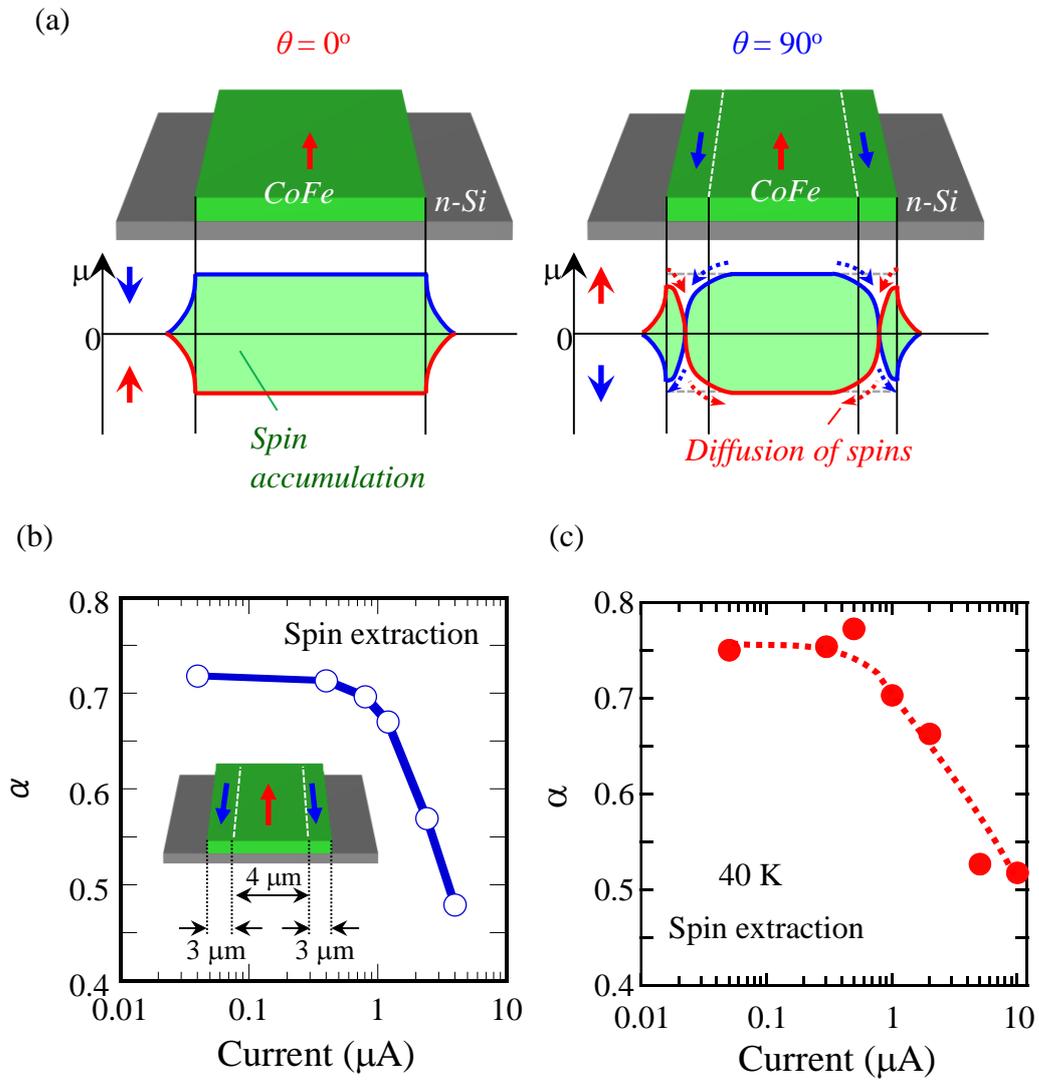

Fig. 3 Y. Ando